# Power Systems Transient Stability Indices: Hierarchical Clustering Based Detection of Coherent Groups Of Generators

Faycal Znidi
Electrical and Computer Engineering Department
Texas A&M University
Texarkana, TX, USA
fznidi@tamut.edu

Hamzeh Davarikia
Electrical and Computer Engineering Department
McNeese State University
Lake Charles, LA, USA
hdavarikia@mcneese.edu

Heena Rathore
Computer Science Department
University of Texas
San Antonio, TX, USA
heena.rathore@ieee.org

*Abstract*— Coherent groups of generators, i.e., machines with perfectly correlated rotor angles, play an important role in power system stability analysis. This paper introduces a real-time methodology based on hierarchical clustering techniques for discovering the degree of coherency among generators using the synchronization coefficient and the correlation coefficient of the generator's rotor angle as the coherency index. Furthermore, the Power Transient Stability Indices (PTSI) were employed to examine the versatile response of the power system. The method uses power systems transients Stability indices, i.e., power Connectivity Factor (CF) index which presents coherently strong generators within the groups, the power Separation Factor (SF) index which unveils to the extent that the generators in different groups tend to swing against the other groups in the event of a disturbance, and the overall system separation index which demonstrates the overall system separation status (CF/SF). The approach is assessed on an IEEE-39 test system with a fully dynamic model. The simulation results presented in this paper demonstrate the efficiency of the proposed approach.

## I. INTRODUCTION

The power network is routinely subjected to a range of disruptions and may become quickly unstable which may lead to a catastrophic blackout because of cascading failure. Coherent groups of generators, i.e., machines with perfectly correlated rotor angles, play an important role in power system stability analysis [1]. The recognition of coherent areas is an important phase to strengthen the power network to avoid potential cascading failures [2]. The aforementioned features can be coordinated along with additional corrective measures to obtain applicable islands in the power system that will help to reduce the complexity of the islanding solution, which is similar to the kind of 0-1 knapsack problem.

Many methods have been introduced in the last decade to solve the classification of coherent areas for operation and control studies. In [3], a general wide-area coherency discovery method based on Project Pursuit (PP) is suggested. Also, a projection cumulative contribution rate (PCCR) index was introduced to establish the central coherency status and the generator coherency groups were formed under different disturbances. Though, the selection of suitable PCCR is yet a problem that requires additional investigation. In [4], the coherent groups of generators were formed using the level of variation of the generator bus voltage phase angles assisted by the hierarchical clustering approach. In [5], a coherency recognition technique uses a linearization of non-equilibrium points is presented. In this methodology, the oscillation status between generators was determined to utilize the eigenvalues of the coefficient matrix of linearized models. In [6], wide area generator speed measurements are employed utilizing Fast Fourier Transform (FFT) based spectral methods. Though, this approach adopts the analysis of linear and stationary data, an assumption that is not always warranted. In [7], generator coherence groups were obtained using the Partitioning Around Medoids (PAM) approach and it was compared with K-means in [8]. The K-means clustering approach is simple to employ and is fast [9], [10]. Nevertheless, the K-means technique has trouble clustering data without the selection of preliminary centroids or the number of clusters and iterations required to determine the clusters [11,12]. In [13], an approach for partitioning the power network based on the versatile response of the system is introduced. In [14], modularity clustering based on community recognition for distribution networks is used to partition the network into coherent islands. In [15, 16], a constrained spectral k-embedded clustering technique is developed to find an islanding solution using the minimal active power flow disruption as the objective function in the meantime addressing the generator's coherency problem. In [17], a two-step Spectral Clustering Controlled Islanding (SCCI) approach is presented at the same time utilizing generator coherency as the only constraint with minimum active power flow disruption to determine a suitable islanding solution.

This paper introduces a new methodology based on a hierarchical clustering technique for identifying the coherency between generators using the correlation index of the rotor angle at the generator to measure the strength of the association between each pair of generators. Further, the transient stability indices are discovered, and the strength of the generators coherency, network integrity, and a complete system separation condition is examined. This method is computationally fast, simple to implement, and does not require a prespecified number of groups. Finally, Simulation results demonstrate the efficiency of the proposed approach.

## II. DYNAMIC COHERENCY DECTTION

### A. Correlation Coefficient as Coherency Measure

Pearson product-moment Correlation Coefficient (CC) is the most widely used correlation to evaluate the degree of the relationship among linearly associated variables. Normally, the correlation coefficient (CC) varies between -1.0 and +1.0 and measures the intensity and direction of the linear relationship that exists among the two quantities. About the power network, the variables represent the coherency of two different generators where a value of +1.0 indicates there is an ideal positive association among the two generators. In order, to evaluate the degree of coherency between two different generators, the CC among generator $i$ and $j$ is formed as follows:

$$CC_{i,j} = \frac{n \sum \delta_i \delta_j - \sum \delta_i \sum \delta_j}{\sqrt{n \sum \delta_i^2 - (\sum \delta_i)^2} \sqrt{n \sum \delta_j^2 - (\sum \delta_j)^2}} \quad (1)$$

where $\delta_i = \{\delta_{i_1}, \delta_{i_2}, \dots, \delta_{i_n}\}$ is the dataset of the rotor angle of the $i^{th}$ generator in the needed time interval including $n$ values of the rotor angle $\delta_i$. Therefore, the coherency between two generators can be revealed by the measurement of the correlation of the rotor angle variation of generators.

### B. Generators Synchronization Coefficients as Coherency Measure

The coherency among the generators, i.e. their tendency to swing together, changes following a disturbance, a characteristic that is used in this methodology to find the generator coherency. Znidi et.al [16] proposed an algebraic model for calculating the synchronization coefficient between $\mathcal{M}$ generators in an $n$ bus system using a classical model for transient analysis. Equation (2) illustrates the synchronization coefficient among generator $i$ and $j$, where $|E_i'|$ is the magnitude of the voltage behind the reactance in the synchronous generator, $B_{ij}$ is the susceptance between the element $i$ and $j$ in the reduced system, and $\delta_{ij}$ is the difference between the rotor angle of machines $i$ and $j$ [17].

$$Ks_{ij} = \sum_{j \in \mathcal{M}}^{m} |E_i'||E_j'|(-B_{ij} \cos \delta_{ij}) \quad (2)$$

There is an associated complete weighted graph $KS_\mathcal{M} = (G, Ks)$ to the power system consisting of $\mathcal{M}$ machine where $G = \{g_1, \dots, g_\mathcal{M}\}$ is the set of generators and $Ks = \{Ks_{ij} | i,j \in G, i \neq j\}$ is the set of synchronization coefficient among the generators. The associated adjacency matrix of the $KS_\mathcal{M}$ is called $KS$ matrix which is a square $\mathcal{M} \times \mathcal{M}$ matrix, that can be easily formed in the real-time fashion for a power network. Hierarchical clustering techniques are employed in the next section to split the generators into the coherent groups of generators.

### C. Hierarchical Clustering

Cluster is the task of grouping a set of objects in the manner that objects in a cluster will be related to one another and separate to the objects in another cluster. Concerning power networks, the clustering technique divides the large network into a variety of lesser ones with a related rotor angle swing curve. In other words, generators in the same cluster will be related to one another and separate from the generators another cluster. The Hierarchical clustering technique is based on the Euclidean distance among all the data tests in the multidimensional space and divides the generators by the distance vector. The number of distances is associated to the similarities among the generators. The number of distances related with the similarities is obtained using the following equation:

$$N_s = \frac{N_g(N_g - 1)}{2} \quad (3)$$

where $N_g$ is the number of generators. Therefore, the fuzzy Hierarchical clustering similarity matrix with a size of $N_g \times N_g$ is used to determine the coherent groups. The components of the similarity matrix are established based on the $KS$ matrix. Figure (1) shows the clustering procedure utilized in the suggested method.

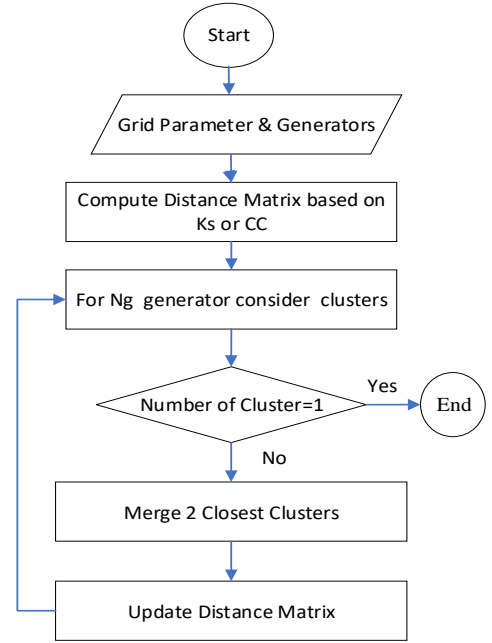

Fig. 1. The Clustering Method

Further, applying the hierarchical clustering techniques, the $KS$ matrix is partitioned to the $\mathcal{N}$ groups of coherent generators. This has an associated $\mathcal{N} \times \mathcal{N}$ matrix called KsGM. Figure (2) shows a generic extraction of KsGM from $KS_{\mathcal{M}}$ for the 10-machine system consisting of three coherent groups of generators. The diagonal elements of matrix KsGM represent the strength of coherency in each group while the off-diagonal elements show how strong is the coherency among different groups. Moreover, the Laplacian matrix and its eigenvalues present valuable information about the power system in real-time. The PTSI based on matrix KsGM is explained in the next section.

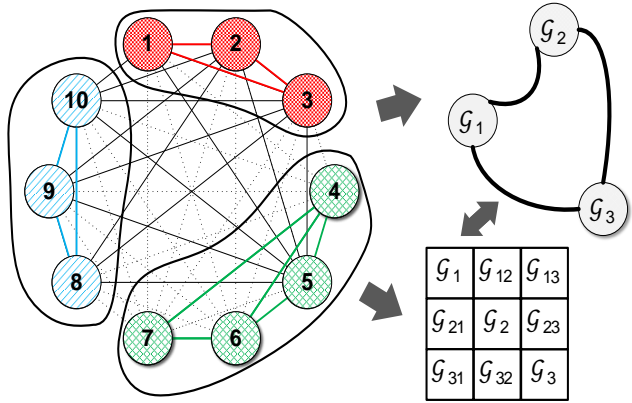

Fig. 2. Formation of Coherent Groups and KsGM from $KS_{\mathcal{M}}$ of 10-machine System

### D. Power Systems Transient Stability Indices

The symmetric square $\mathcal{N} \times \mathcal{N}$ matrix $KsGM = (KsGM_{ij} | i, j \in \mathcal{N})$ is the adjacency matrix associated with the complete graph of $\mathcal{N}$ coherent groups of generators which has interesting properties that can be extracted by applying a simple algebraic process. The diagonal and off-diagonal elements present the strength of coherency between the generators inside a group and between the groups, respectively. Acquiring matrix KsGM in real-time fashion paws the way for future analysis to observe how is the integrity of the power network. To this end, the next subsections define the PTSI based on matrix KsGM.

The power Connectivity Factor (CF) index, is called as the mean of diagonal of the KsGM matrix, which presents coherently strong generators within the groups. The power Separation Factor (SF) index is called the mean of KsGM matrix off-diagonal which unveils to the extent that the generators in different groups tend to swing against the other groups after a disturbance. The overall system separation status (CF/SF) is defined as CF divided by SF, which shows the total system splitting status. Eq. 4, Eq. 5, and Eq. (6) demonstrate CF, SF, and the CF/SF respectively, where $a_{ii}$ is the diagonal element of $KsGM$.

$$CF_i = a_{ii} \quad (4)$$

$$SF_{ij} = \sum_{i=1}^{n-1} \sum_{j=2}^{n} \frac{a_{ii} + a_{ij}}{2 \times a_{ij}} \quad (5)$$

$$CF/SF = \frac{\sum_{i=1}^{n} a_{ii}}{\sum_{i=1}^{n-1} \sum_{j=2}^{n} a_{ij}} \quad (6)$$

In this paper, the measure for determining the generators of one cluster is based on the data of the similarity matrix of the Ks and CC.

### III. SIMULATION TEST CASES

The methodology efficiency is assessed via the simulation study performed on the modified IEEE 39-bus system shown in Fig. 3. The approach has been executed in MATLAB and all time-domain simulations are attained in DIgSILENT PowerFactory. Table 1 list the events that occurred as a result. Using the proposed approach, the generator coherence groups will be identified for different fault locations.

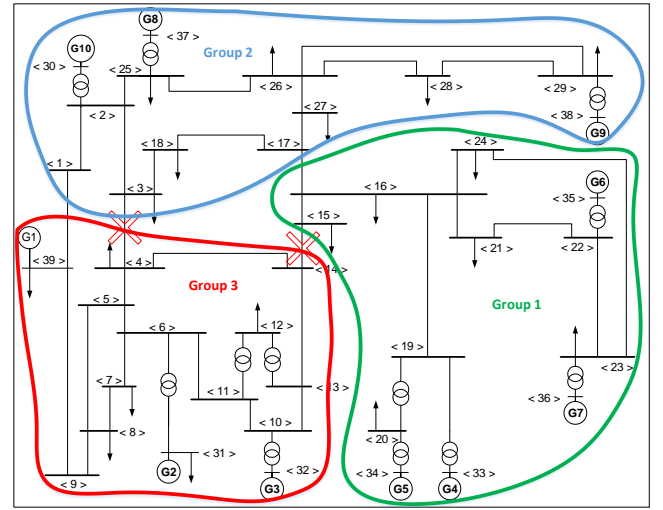

Fig. 3. EEE 39 bus system. The crosses are the event in the sample case study in scenario 1

### A. Scenario 1: Stable Case

Table 1 list the events that occurred as a result of the duration of the simulation time of 100 s.

Table 1. Events Occurred in Scenario 1

| Time (s) | Description |
|---|---|
| 2.00 | Short circuit on lines 3-4 |
| 2.40 | Switch event on lines 3-4 |
| 20.00 | Short circuit on lines 14-15 |
| 20.40 | Switch event on lines 14-15 |

The outage of lines 3-4 and 14-15 at t=2 and t=20 s respectively are considered as the events. Figures (4) and (5) show the generator rotor angle oscillation and the bus frequencies respectively. As can be seen, the rotor angle fluctuations are damped, and all the generators stay in synchronization while groups of generators develop into stronger after the events. Figure (4) shows the stability of the power system following the occurrence of the disturbances.

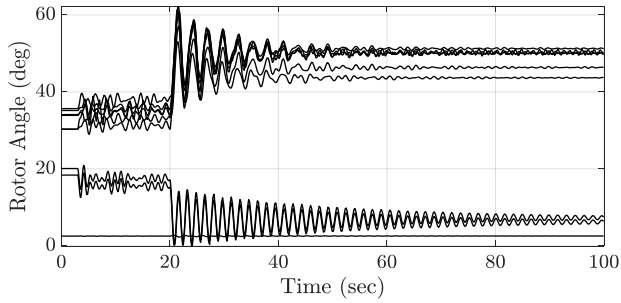

Fig. 4. Generator rotor angles in scenario 1

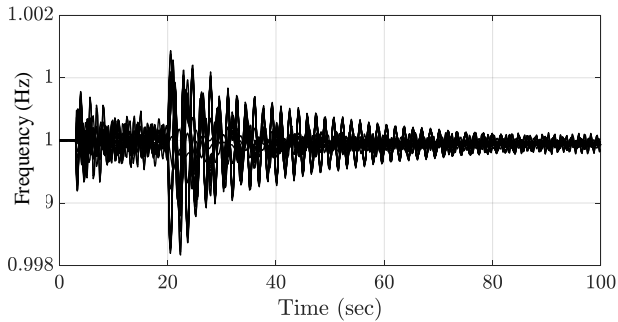

Fig. 5. Bus frequencies in scenario 1

Therefore, by using the similarity matrix and utilizing the hierarchical clustering technique, the coherent groups of generators of the power network can be recognized. As seen in Figure (3), grid generators are grouped into three separate groups: $\{G_4, G_5, G_6, G_7\}$; $\{G_8, G_9, G_{10}\}$; $\{G_1, G_2, G_3\}$. After the second outage, the CC between generators is calculated from Equation (1) and the CC matrix is formulated. Applying hierarchical clustering on the CC results in three separate coherent groups of generators shown in Figure (3). The power transient stability indices CF, SF, and (CF/SF) as shown in Figures (6), (7) and (8) respectively become steady at 90 s.

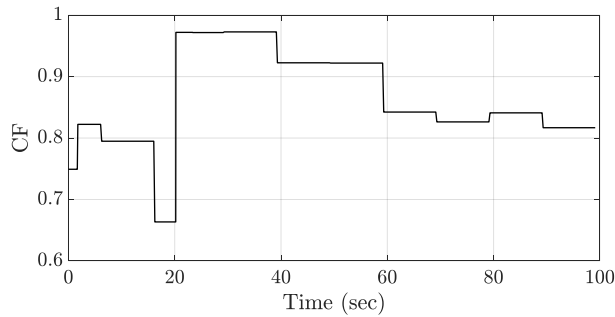

Fig. 6. The behavior of the power Connectivity Factor (CF) index using CC in scenario 1

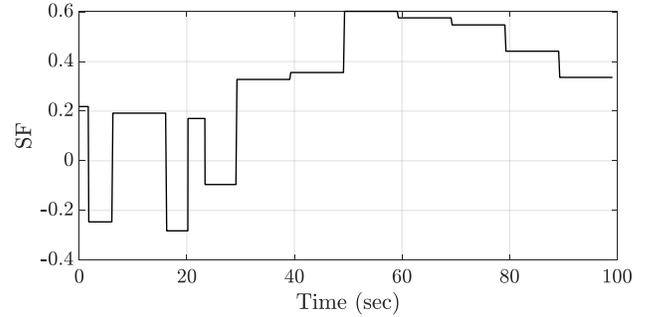

Fig. 7. The behavior of the power Separation Factor (SF) index using CC in scenario 1

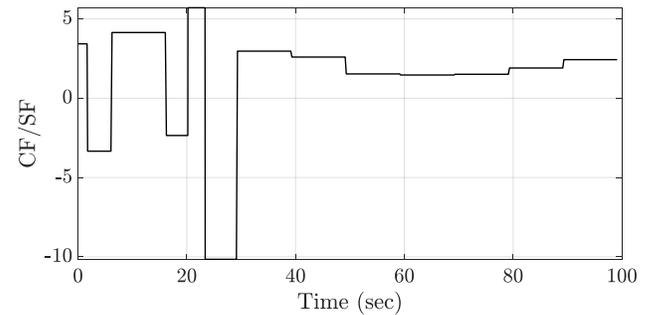

Fig. 8. The behavior of the overall system separation status (CF/SF) using CC in scenario 1

As can be seen, after the first event at t=2s the CF and SF are slightly increased because the events cause the generators to swing, however, this failure did not cause total separation. The CF and SF have small variations because the generators stabilize between t=2 s and t=20 s. After the second event at t=20 s, the CF and SF are increased and decreased respectively because the faults cause the group of coherent generators to be separated.

Similarly, the Ks among generators is calculated from (2) and the Ks matrix is formulated. Applying hierarchical clustering on the Ks results in three separate coherent groups of generators shown in Figure (4). The power transient stability indices CF, SF, and (CF/SF) are shown in Figures (9), (10), and (11) respectively become steady at 40 s.

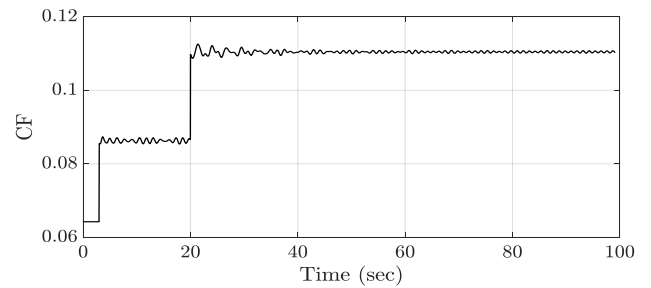

Fig. 9. The behavior of the power Connectivity Factor (CF) index using Ks in scenario 1

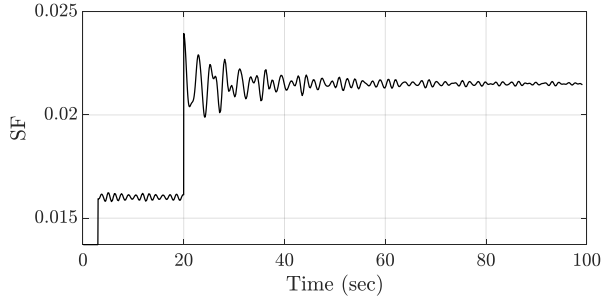

Fig. 10. The behavior of the power Separation Factor (SF) index using Ks in scenario

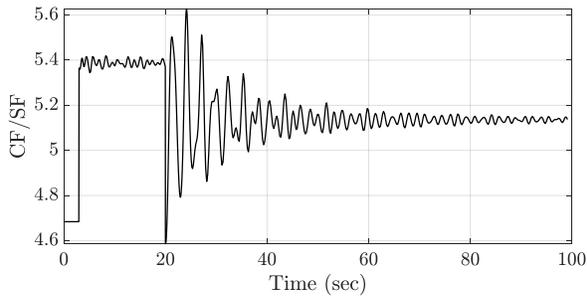

Fig. 11. The behavior of the overall system separation status (CF/SF) using Ks in scenario 1

### B. Scenario 2: Unstable Case

Three short circuits (SC) events occurred in lines 3-4 at t=3 s and lines 13-14 and 16-17 at t= 10 s. Table 2 list the events that occur in the testbed system during the simulation time of 30 s.

Table 2. Events Occurred in Scenario 2

| Time (s) | Description |
|---|---|
| 3.00 | Short circuit on lines 3-4 |
| 10.00 | Short circuit on lines 13-14 and 16-17 |
| 15.20 | Switch event |

Figures (12) and (13) demonstrate the generator's rotor angle and the system frequency respectively, which indicates the system instability following the second event.

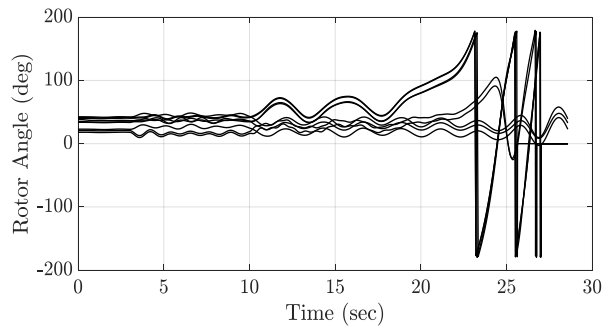

Fig. 12. Generators rotor angles in scenario 2

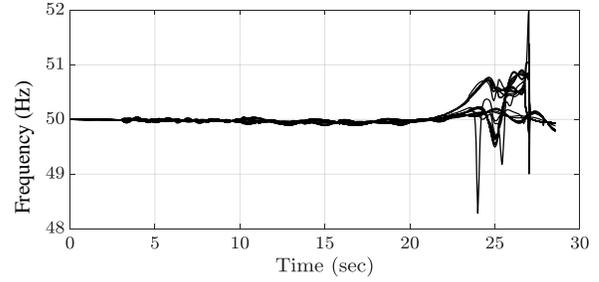

Fig. 13. Bus frequencies in scenario 2

Using the CC among the generator as coherency measure and applying the hierarchical clustering methodology results in the variation of the CF, SF and CF/SF as shown in Figures (14), (15) and (16) respectively following the events. As can be seen, after the first event at t=3s the CF and SF are slightly increased because the events cause the generators to swing, however, this failure causes total separation at 25.25 s. The CF and SF have small variations because the generators stabilize between t=3 s and t=25 s. After the second event at t=20.25 s, the CF and SF become unstable. The behavior of the overall system separation status (CF/SF) using CC is shown in Figure (16).

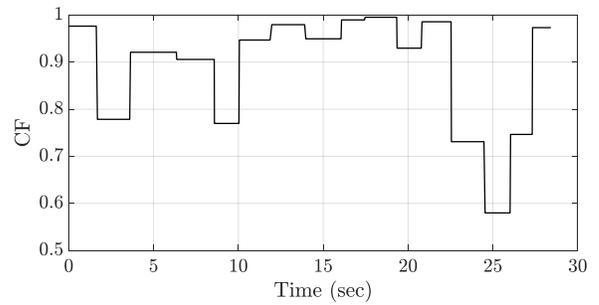

Fig. 14. The behavior of the power Connectivity Factor (CF) index using CC in scenario 2

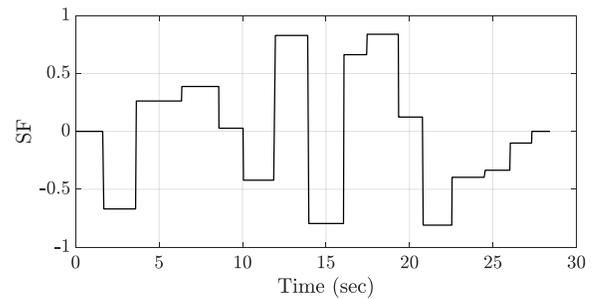

Fig. 15. The behavior of the power Separation Factor (SF) index using CC in scenario 2

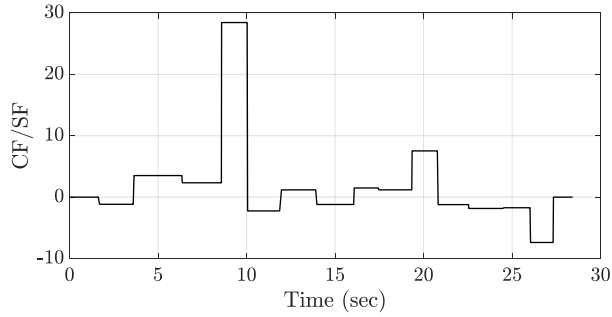

Fig. 16. The behavior of the overall system separation status (CF/SF) using CC in scenario 2

Similarly, the Ks among generators are calculated from Equation (2) and the Ks matrix is formulated. Applying hierarchical clustering on the Ks results in the variation of the CF, SF and CF/SF shown in Figures (17), (18) and (19) respectively following the events. The power transient stability indices CF, SF, and (CF/SF) become steady at 40 s. The CF, SF and CF/SF have small variations because the generators stabilize between t=3 s and t=25 s. After the second event at t=20.25 s, the CF and SF become unstable.

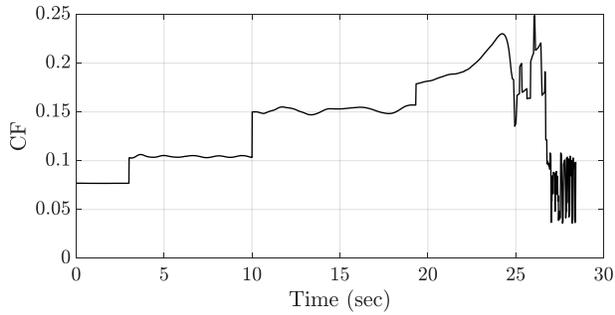

Fig. 17. The behavior of the power Connectivity Factor (CF) index using Ks in scenario 2

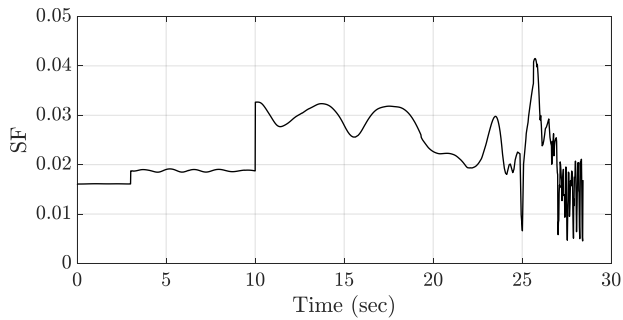

Fig. 18. The behavior of the power Separation Factor (SF) index using Ks in scenario 2

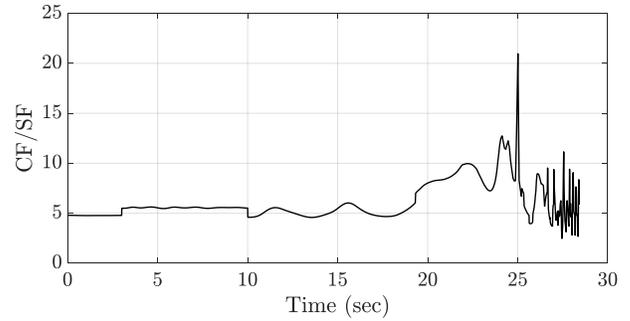

Fig. 19. The behavior of the overall system separation status (CF/SF) using Ks in scenario 2

IV. CONCLUSION

This paper proposes a novel methodology for discovering the degree of coherency among generators. It uses the synchronization coefficient and the correlation coefficient to measure the strength of the association between each pair of generators. The hierarchical clustering techniques were used to find the generator's coherency. Further, the strength of the generators coherency was assessed, the power systems transient stability indices, the integrity indices, and the overall system status was examined. It was evident from the results that this approach can determine the degree of coherency among any pair of generators accurately using synchronization coefficient and the correlation coefficient among the generators.

REFERENCES

[1] ZNIDI, F., DAVARIKIA, H., IQBAL, K. et al. Multi-layer spectral clustering approach to intentional islanding in bulk power systems. J. Mod. Power Syst. Clean Energy 7, 1044–1055 (2019) doi:10.1007/s40565-019-0554-1

[2] H. Davarikia, F. Znidi, M. R. Aghamohammadi, and K. Iqbal, "Identification of coherent groups of generators based on synchronization coefficient," in Power and Energy Society General Meeting (PESGM), 2016, 2016, pp. 1-5: IEEE.

[3] T. Jiang et al., "Projection pursuit: A general methodology of wide-area coherency detection in bulk power grid," in *IEEE Transactions on Power Systems*, vol. 31, no. 4, pp. 2776-2786, 2016.

[4] Ali, M.et al., "Detection of coherent groups of generators and the need for system separation using synchrophasor data", in Proc. *Int. Conf. on Power Engineering and Optimization*, pp. 7–12, 2013.

[5] Ma, Zhenbin et al., "The Application of a Generator Coherency Identification Method Based on Linearization in Complex Power System.", in Proc. *China International Conference on Electricity Distribution (CICED)*, August 10-13, 2016.

[6] M. Jonsson, M. Begovic, and J. Daalder, "A new method suitable for real-time generator coherency determination," in *IEEE Trans. Power Syst.*, vol. 19, no. 3, pp. 1473–1482, 2004.

[7] Pyo, G.C., Park, J.W., and Moon, S.I. "A new method for dynamic reduction of power system using PAM algorithm," in Proc. *IEEE Power and Energy Society General Meeting*, 2010.

[8] S. K. Joo et al., "Coherency and aggregation techniques incorporating rotor and voltage dynamics", in *IEEE Trans. on Power Systems.*, vol. 19, no. 2, pp.1068-1075, 2004.

[9] Rathore, H. et al., "Novel approach for security in wireless sensor network using bio-inspirations". In Proc. *Sixth International Conference on Communication Systems and Networks (COMSNETS)* (pp. 1-8), 2014.

[10] Rathore, H. and Jha, S., "Bio-inspired machine learning-based wireless sensor network security". In Proc. *World Congress on Nature and Biologically Inspired Computing (pp. 140-146)*, 2013.

[11] Kalpana D. Joshi et al., "Modified K-Means for Better Initial Cluster Centres", *International Journal of Computer Science and Mobile Computing, IJCSMC*, vol. 2, no. 7, pg. 219– 223, 2013.

[12] M. H. R. Koochi, S. Esmaeili, and G. Ledwich, "Taxonomy of coherency detection and coherency-based methods for generators grouping and power system partitioning," in *IET Generation, Transmission & Distribution*, vol. 13, no. 12, pp. 2597-2610, 2019.

[13] M. Rezaeian Koochi, S. Esmaeili, and P. Dehghanian, "Coherency detection and network partitioning supported by the wide-area measurement system," in Proc. *IEEE Texas Power Energy Conf.*, pp. 1–6, 2018.

[14] H. Zhang, M. Peng and P. Palensky, "Intentional islanding method based on community detection for distribution networks," in *IET Generation, Transmission & Distribution*, vol. 13, no. 1, pp. 30-36, 8 1 2019. DOI: 10.1049/iet-Ltd.2018.5465, 2019.

[15] Ahad E., and Mladen K., "Controlled Islanding to Prevent Cascade Outages Using Constrained Spectral k-Embedded Clustering", in *Power Systems Computation*, 2016.

[16] Quirós-Tortós, J. et al., "Determination of Sectionalising Strategies for Parallel Power System Restoration: A Spectral Clustering-Based Methodology", *Electric Power Systems Research*, 116, pp. 381-390, 2014.

[17] L. Ding et al., "Two-Step Spectral Clustering Controlled Islanding Algorithm", in *IEEE Transactions on Power Systems*, vol. 28, no. 1, pp. 75-84, 2013.

[18] F. Znidi, H. Davarikia, and K. Iqbal, "Modularity clustering-based detection of coherent groups of generators with generator integrity indices," in *IEEE Power & Energy Society General Meeting*, pp. 1-5, 2017.

[19] H. Davarikia, M. Barati, F. Znidi, and K. Iqbal, "Real-Time Integrity Indices in Power Grid: A Synchronization Coefficient Based Clustering Approach," in Power & Energy Society General Meeting, 2018 IEEE, 2018, pp. 1-5: IEEE